\documentclass[conference]{IEEEtran}
\IEEEoverridecommandlockouts

\usepackage{hyperref}
\usepackage{cite}
\usepackage{amsmath,amssymb,amsfonts}
\usepackage{algorithmic}
\usepackage{graphicx}
\usepackage{textcomp}
\usepackage{xcolor}

\def\BibTeX{{\rm B\kern-.05em{\sc i\kern-.025em b}\kern-.08em
    T\kern-.1667em\lower.7ex\hbox{E}\kern-.125emX}}

\begin{document}
\title{Self-supervised Multi-Modal Video Forgery Attack Detection
%\\
}

\author{
\IEEEauthorblockN{Chenhui Zhao\IEEEauthorrefmark{1} and
Xiang Li\IEEEauthorrefmark{2} 
% and Siwen Dong\IEEEauthorrefmark{3} 
and Rabih Younes\IEEEauthorrefmark{3}}

\IEEEauthorblockA{
\IEEEauthorrefmark{1}Department of Electrical and Computer Engineering, University of Michigan, Ann Arbor, USA\\
\IEEEauthorrefmark{2}Department of Electrical and Computer Engineering, Carnegia Mellon University, Pittsburgh, USA\\
\IEEEauthorrefmark{3}Department of Electrical and Computer Engineering, Duke University, Durham, USA
}}

% \author{
% \IEEEauthorblockN{1\textsuperscript{st} Chenhui Zhao}
% \IEEEauthorblockA{\textit{dept. Electrical and Computer Engineering} \\
% \textit{University of Michigan}}
% \and
% \IEEEauthorblockN{2\textsuperscript{nd} Xiang Li}
% \IEEEauthorblockA{\textit{dept. Electrical and Computer Engineering} \\
% \textit{Carnegie Mellon University}}
% \and
% \IEEEauthorblockN{3\textsuperscript{rd} Siwen Dong}
% \IEEEauthorblockA{\textit{dept. Electrical and Computer Engineering} \\
% \textit{Duke University}}
% \and
% \IEEEauthorblockN{4\textsuperscript{th} Rabih Younes}
% \IEEEauthorblockA{\textit{dept. Electrical and Computer Engineering} \\
% \textit{Duke University}}
% \and
% \and
\maketitle

\begin{abstract}
\textbf{
  Video forgery attacks threaten surveillance systems by replacing the video captures with unrealistic synthesis, which can be powered by the latest augmented reality and virtual reality technologies. From the machine perception aspect, visual objects often have RF signatures that are naturally synchronized with them during recording. In contrast to video captures, the RF signatures are more difficult to attack given their concealed and ubiquitous nature. In this work, we investigate multimodal video forgery attack detection methods using both visual and wireless modalities. Since wireless signal-based human perception is environmentally sensitive, we propose a self-supervised training strategy to enable the system to work without external annotation and thus adapt to different environments. Our method achieves a perfect human detection accuracy and a high forgery attack detection accuracy of 94.38\% which is comparable with supervised methods. The code is publicly available at: \href{https://github.com/ChuiZhao/Secure-Mask.git}{https://github.com/ChuiZhao/Secure-Mask.git}
  }
\end{abstract}


\section{Introduction}

In recent years, the unique properties e.g. concealment, penetration, and ubiquity have been extensively investigated in wireless-based perception methods. Person-in-WiFi \cite{person-in-wifi} is a pioneering work attempting to address fine-grained human perception problems by using WiFi signals.  The follow-up works Secure-Pose \cite{huang2021crossmodal} and its improved version \cite{huang2021forgery} propose learning-based methods to detect video forgery attacks using radio-frequency (RF) signals. Besides the video forgery attack task, RF-based methods can also achieve comparable performance against visual-based methods in other visual representation tasks \cite{person-in-wifi, WiPose, Wi-Pose}. However, most RF-based human perception methods are environmentally sensitive \cite{adib2013see, huang2014feasibility, person-in-wifi, Wi-Pose, yang2020temporal}, thus it is hard to adapt well-trained models to unseen environments, which severely prevents them from practical applications. 

To date, several methods have been tried to tackle the adaptation problem. Person-in-WiFi \cite{person-in-wifi} proposes a style-transfer method for CSI measurements utilizing the Cycle-GAN \cite{zhu2017unpaired}, but the performance gain of the proposed module is limited even after complicated synthetic data generation. WiPose \cite{WiPose} extracted the environmental weak-dependent Body-coordinate velocity profile(BVP) from CSI measurement combing with an antenna selection strategy to ease the influence of the background environment. Although many attempts have been made to mitigate the impact of environment changes \cite{person-in-wifi, WiPose, Wi-Pose}, the cross-environment adaptation remain unfeasible since the RF-based human perception heavily depends on the Doppler effect and electromagnetic property differences between human and background environment. 
Consequently, the gaps between RF-based human perception performance in the seen and unseen environment are hard to bridge from the perspective of data augmentation and denoising.

On the other hand, a surveillance camera is a periphery device for surveillance systems which captures visual information of target environments and transmits it to a central server. Due to the redundancy of the video modality, the video compression is always conducted before transmission. Motion vector is one of the essential components of the compressed video which reflects the block-wise spatial position changes across frames, and \textit{de facto} fits the format of the label of the RF-based perception system.

\begin{figure}[t]
    \centering
	\includegraphics[width=\linewidth]{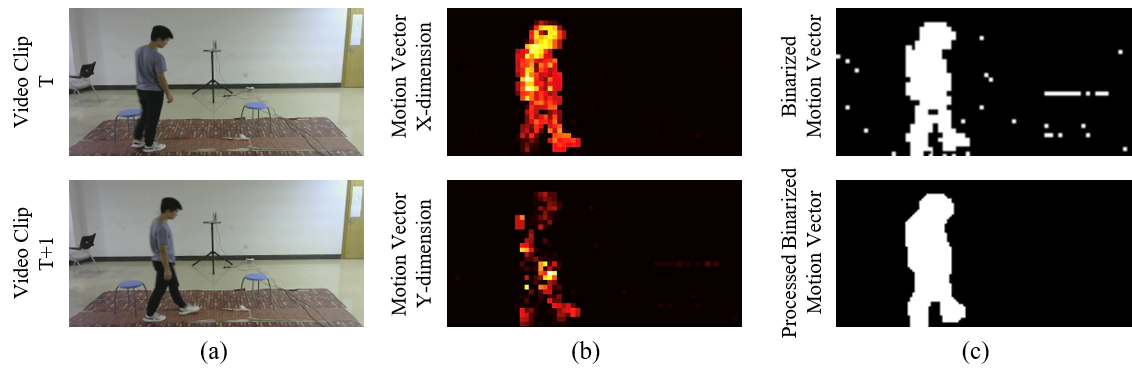}
    \vspace{-0.8cm}
    \caption{Illustration of an original 2D motion vector and a binarized motion vector produced by our method from compressed video. The top and bottom figures of (a) are two adjacent frames in the video. The top and bottom figures of (b) are the X-dimension motion vector and the Y-dimension motion vector. The top and bottom figures of (c) are the original binarized motion vector and the binarized motion vector after processing.}
\label{figure1}
\vspace{-0.3cm}
\end{figure}

In this work, to solve the adaptation problem of RF-based perception method, we introduce a self-supervised learning scheme to enable the model to learn from compressed video streams and further leverage it to conduct video forgery detection. We name the proposed system Secure-Mask. In particular, we adopt motion vectors from compressed video streams to create the supervision for RF-based model. Fig.~\ref{figure1} shows an example of motion vector and generated mask. Compared with other frame-based video forgery detection methods \cite{hyun2013forgery, fadl2021cnn}, Secure-Mask can work in a real-time manner to generate fine-grained human segmentation and detect video replacement attacks at the object level. The concealed and ubiquitous properties of WiFi signals make it a good alternative to surveillance video and can be suitable for future secure systems to act on when cameras are offline, occluded or attacked. Moreover, the self-supervised training scheme enables Secure-Mask to adapt to the new environment which eliminates the redundant data labeling after environmental changes (i.e. furniture movement). The contributions of this paper are as follows. 
\begin{itemize}
    \item 
    We propose a self-supervised learning scheme for RF-based human perceptions leveraging the motion vector as a source of supervision and proving its ability to work without external annotations.
    \item 
    We built up a self-supervised video forgery attack detection pipeline that can act in a real-time manner with high accuracy of 94.38\%. The performance of Secure-Mask is comparable to its supervised counterpart Secure-Pose \cite{huang2021crossmodal}.
\end{itemize}

\section{Related Works}

\noindent\textbf{Camera-based human perception.}\quad In the computer vision field, many works \cite{wang2020solo, Mask-RCNN} used well-developed feature extraction methods to accomplish challenging human perception tasks. In addition, there have been many works in object segmentation \cite{li2022hybrid, li2022video,li2022online,li2022r,li2022panoramic}, pose estimation \cite{sun2019deep, martinez2017simple}, and activity recognition \cite{ijjina2017human, garcia2018modality}. More recently, depth cues obtained by the RGB-D camera have been introduced in human perception tasks and some other works \cite{palmero2016multi, zhou2017improving} have shown that depth cure can improve performance.

\noindent\textbf{Sensor-based human perception.}\quad 
The Frequency Modulated Continuous Wave (FMCW) radar system was first introduced by Adib et al. \cite{adib2015capturing} to capture coarse human bodies with a delicate radar device. Later, they extended this system to do pose estimation through the wall or other occlusions, with 2D \cite{zhao2018through} and 3D \cite{zhao2018rf} included. 
Compared with the above methods, which rely heavily on expensive Radar equipment, WiFi signals provide a more ubiquitous and cheap option. However, WiFi-based works \cite{adib2013see, huang2014feasibility} were not popularized before because they have not been producing fine-grained human masks or human skeletons until Wang et al. proposed Person-in-WiFi \cite{person-in-wifi}. After that, more and more researchers paid attention to WiFi-based human perception works. For example, WiTA \cite{yang2020temporal} recognized human activity in an attention-based way using commercial WiFi devices. Wi-Pose \cite{Wi-Pose} reconstructed fine-grained human poses using WiFi signals. In addition, some previous works \cite{li2020activitygan,luo2021toward} also investigate how to augment data to achieve better performance.

\noindent\textbf{Video Forgery Detection.}\quad Forgery detection for surveillance systems has drawn researchers' attention due to the advanced video forgery technologies. Many researchers solved this problem by analyzing the spatiotemporal features in surveillance video \cite{hyun2013forgery, fadl2021cnn}. These methods can determine the frame-based forgery, for example, frame delete and insert. For the object-based video forgery, Mohammed et al. proposed a sequential and patch analysis method \cite{9089005}, which can generate coarse forgery traces in each surveillance video frame. Relatedly, SurFi \cite{lakshmanan2019surfi} compared timing information from WiFi signals and the corresponding live video to detect camera looping attacks. To generate fine-grained forgery traces while detection, Secure-Pose \cite{huang2021crossmodal} first proposed a cross-modal system that can detect and localize forgery attacks in each video frame through a supervised way.

\section{Secure-Mask System}

In this section, we will elaborate on the detailed pipeline of our proposed Secure-Mask system.

\subsection{System Overview}

\begin{figure}[t]
	\centering
	\includegraphics[width=\linewidth]{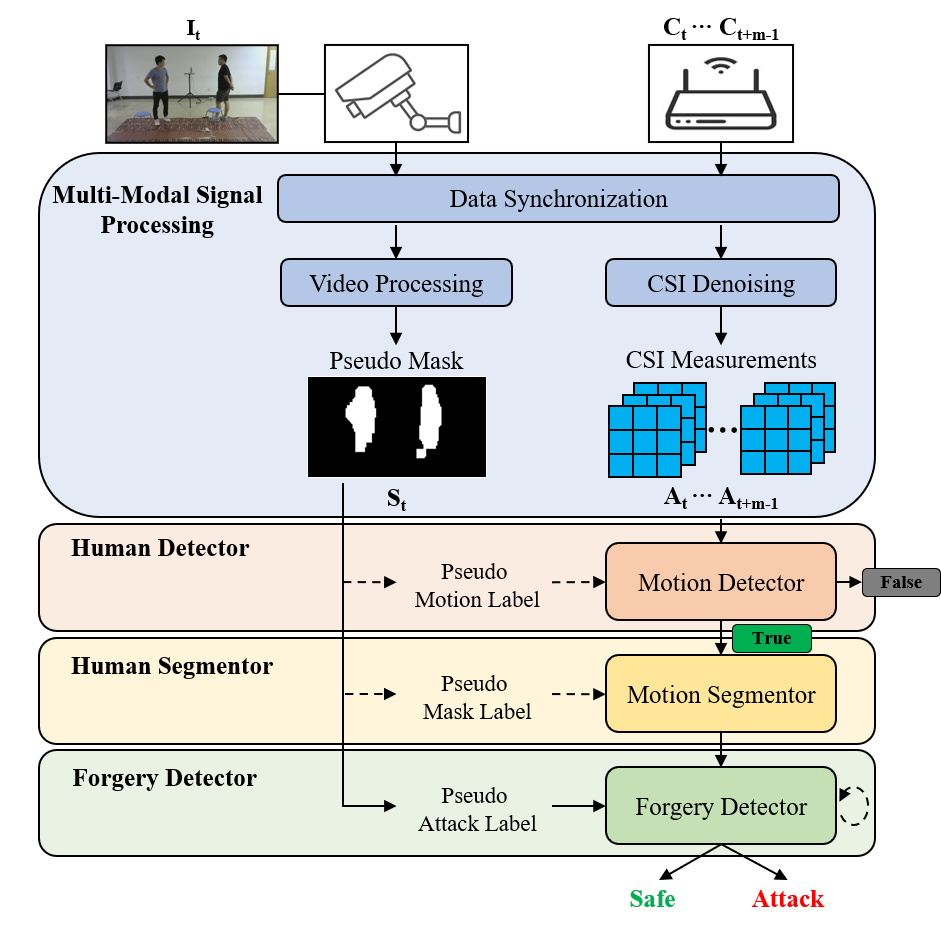}
    \vspace{-0.8cm}
	\caption{Secure-Mask overview. Dash lines indicate model updating. Secure-Mask can be boiled down to four modules: multi-modal signal processing, human detector, human segmentor, and forgery detector. In the training phase, the motion vector $M_t$ of video frame $I_t$ is first synchronized to $m$ CSI measurements $\{C_t,\cdots,C_{t+m}\}$. Then we conduct the preprocessing separately for video and CSI data to get binarized masks $S_t$ and CSI measurements in matrix form $\{A_t, \cdots, A_{t+m}\}$. After that, we generate annotations for training the human detector and human segmentor networks. Finally, we use the predicted masks acquired from the human segmentor to generate annotations for training the forgery detector.
	%\ea{Add the notation of motion vector and CSI measurements explicitly in the figure.}
	}
	\label{figure2}
\end{figure}

The Secure-Mask is composed of four parts: multi-modal signal processing, human detector, human segmentor, and forgery detector. As shown in Fig.~\ref{figure2}, in the training phase, we leverage the motion vector in the live video stream to update the networks. In the multi-modal signal processing module, we first extract the motion vector $\{M_t\}_{t=0}^{T}$ from video frames $\{I_t\}_{t=0}^T$ and synchronize it with the CSI measurements $\{C_t\}_{t=0}^{T}$, then conduct signal processing separately. After that, the processed masks serve as labels to update networks. In the inference phase, only the CSI branch in the multi-modal signal processing module is activated. The processed CSI measurements are sent to the human detector to detect motions. To improve the efficiency of the whole system, the human segmentor and forgery detector only acts when the human detector confirms moving objects.

\subsection{Multi-Modal Signal Processing}

Secure-Mask is a cross-modal system that takes advantage of both visual and wireless modalities. To ensure both modalities contain homogeneous information, the data synchronization is essential. Given the nature of wireless communication, wireless signals always have a much higher sample rate than video frames. Therefore, we assign multiple wireless frames to one video frame. Let us denote the captured video frames as $\{I_t\}_{t=0}^T$ and CSI measurements $\{C_t\}_{t=0}^T$ where $I_t\in\mathbb{R}^{3\times H\times W}$ and $C_t\in\mathbb{C}^{K\times N_{tx}\times N_{rx}}$. $H$ and $W$ are the height and width of the video frames. $N_{tx}$, $N_{rx}$ and $K$ are the number of transmitters, receivers and subcarriers, respectively. We assign $m$ CSI measurements $\{C_t,\cdots,C_{t+m-1}\}$ to one frame $I_t$. We consider the amplitude of CSI measurements $\{A_t\}_{t=0}^T$ to make it a real matrix. After that we get the data pairs $\{I_t, \{A_t,\cdots,A_{t+m}\}\}_{t=0}^{T}$.

\noindent\textbf{CSI Processing.}\quad The environment noise can cause sudden changes in the CSI measurements, which will impact the efficiency of extracting the amplitude features from it. To filter out outliers in the CSI measurements, we utilize the Hampel identifier \cite{hampel-identifier} to denoise the CSI data as Secure-Pose \cite{huang2021crossmodal} did.

\noindent\textbf{Video Processing.}\quad To improve video storage and transmission, it is common to perform video compression. Typically, the compression techniques such as MPEG-4 and H.264 leverage the temporal continuity of successive frames and retain only a few complete frames while reconstructing other frames using the motion vector and residual error. Our solution utilizes the 2D motion vector to create masks, which can be separated into binarization, denoising, and refinement. The motion vectors within a group of pictures (GOP) can be denoted as $\{M_t\}_{t=0}^G$ where $M\in\mathbb{R}^{H\times W\times 2}$. Since the velocity of human activity is slow compared to the video frame rate, by using a short GOP length, the human movement can be assumed as the same in each GOP. We determine the binary mask of the human movement from two dimensions of the motion vector, angle and amplitude. Let $\hat{M}$ be the sum of all motion vectors in a GOP and $\bar{M}$ be the $\phi_{3\times 3}(\hat{M})$ where $\phi_{3\times 3}$ is an gaussian smooth function with a kernel size of 3. For a single GOP, the binarized mask can be denoted as $S$ and computed as:
\begin{equation}
	\label{binarization}
	S=\left\{
		\begin{aligned}
			1, \quad&if \quad \Vert \hat{M}_{i,j} \Vert_2 + \lambda \frac{<\bar{M}_{i,j},\hat{M}_{i,j}>}{|\bar{M}_{i,j}|\cdot|\hat{M}_{i,j}|} \geqslant \tau \\
			0, \quad&else
		\end{aligned}
	\right.
\end{equation}
where $\lambda$ and $\tau$ are constants. The first term and second term filter motion vector based on amplitude and degree respectively. Here we set $\lambda = 1$ and $\tau = 0.5$. The mask $S$ is further processed through a stack of soothing and morphological operations after binarization.

\subsection{Human Detector}
The human detector network is a lightweight network that aims to determine the existence of human motion. The Human detector takes CSI as input and outputs the binary result judging the human movement. We utilize both the convolution layer and the Long-Short Term Memory (LSTM) layer to process the CSI data, as it includes both spatial and temporal features. In particular, the CSI measurements $\{A_t,\cdots, A_{t+m}\}$ are concatenated in the subcarriers dimension to form $A^c_t\in \mathbb{R}^{mK\times N_{tx}\times N_{rx}}$. The CSI data $A^c_t$ is first processed by two convolution layers followed by one LSTM layer. The final output is obtained by applying two linear layers to the LSTM output. With the human detector, we can save computational resources by only activating the following modules after detecting human motion. 

We supervise the human detector by binary cross-entropy loss. Since the human detector is a binary classification network, the binarized ground-truth is obtained by the following criterion.
\begin{equation}
\label{criterion motion}
\mathcal{C}(S)=\left\{
\begin{aligned}
1 \qquad&if \qquad \frac{\sum^W_{i=0}\sum^H_{j=0}S_{i,j}}{W\times H}\geq\eta\\
0  \qquad&else
\end{aligned} 
\right. 
\end{equation}
where $W$ and $H$ are the width and height of the pseudo mask $S$ and $\eta$ is a constant threshold, and we set $\eta=0$.

\subsection{Human Segmentor}

Given the concatenated CSI data $A^c_t$, the human segmentor generates masks of moving humans in the perception field. To conduct this challenging task, an UNet-liked structure is leveraged. The input CSI data $A^c_t$ is first tiled to image size before feeding to the network. After that, the upsampled tensor is fed into an encoder to produce the encoded feature map. Then a transposed convolution-based decoder is utilized to transform wireless features to image space. Both encoder and decoder contain four downsample and upsample operations with a stride of 2 and, after each scaling operation, a 2D convolution is involved to refine the feature map before the next scaling. In particular, skip connections are used to retain swallow wireless features in later layers. Let us denote the human segmentor as $\mathcal{S}$, the human mask prediction $P_t\in \mathbb{R}^{H\times W}$ can be denoted as $P_t=\mathcal{S}(A^c_t)$

We supervise the human segmentor by binary cross-entropy loss $\mathcal{L}_{bce}$ and Dice loss $\mathcal{L}_{Dice}$. The overall loss for training is $\mathcal{L}=\mathcal{L}_{Dice} + \lambda_{b}\mathcal{L}_{bce}$.

\subsection{Forgery Detector}
The forgery detector aims to detect video forgery attacks using multimodal data. Since wireless perception heavily depends on the Doppler effect which is caused by the motion of objects, the motion vector and wireless data contain homogeneous representations of the moving humans in the perception area. We leverage the human mask as a proxy to conduct contrastive learning for video forgery detection. Given a clip of video $\{I_t,\cdots,I_{t+g}\}$, we can obtain the motion vectors $\{M_t,\cdots,M_{t+g}\}$ freely from the compressed video streams. We further processed it to obtain the pseudo mask $\{S_t,\cdots,S_{t+g}\}$. The human segmentor predicts the human masks $\{P_t,\cdots,P_{t+g}\}$ from the wireless modality. We tailor a network to compare the human masks from visual and wireless modalities to determine their homogeneity. The masks from the visual modality $\{S_t,\cdots,S_{t+g}\}$ and masks from the wireless modality $\{P_t,\cdots,P_{t+g}\}$ are concatenated as $\{Q_t,\cdots,Q_{t+g}\}$ where $Q_t\in \mathbb{R}^{2\times H\times W}$. We extract features for each time step separately with a ResNet-based network. After that, extracted features are fed into a one-layer LSTM followed by two fully connected layers before the final output. We predict the homogeneity in a clip-wise manner. 

Similar to the human detector, we utilize binary cross-entropy to supervise the training. We generate unmatched input sequence pairs to synthesis video forgery attacks by selecting unsynchronized masks pairs.

\section{Experiments}

To the best of our knowledge, there is no public multi-modal dataset for video forgery. Thus, we conduct experiments using the same dataset used in Secure-Pose \cite{huang2021crossmodal}. In this section, we will show the qualitative result for the human segmentor and the quantitative result for both the human detector and forgery detector. In the quantitative result, we report the evaluation metrics including the accuracy (Acc), the false positive rate (FPR), and the true positive rate (TPR).

\subsection{Dataset Description}
The dataset (same as the dataset used in \cite{huang2021crossmodal}) was collected in an $8m\times 16m$ office room with 5 volunteers. As shown in Table~\ref{table1}, during the experiment phase, zero to three volunteers were asked to perform walking, sitting, waving hands, or random movements concurrently in the perception area.

\begin{table}[htb]
	\caption{Statistic of the dataset. P: Number of concurrent person. F: number of video frames.}
	\begin{center}
	\begin{tabular}{cccccc}
		\hline\hline
		P & 0 & 1 & 2 & 3 & total\\
		\hline
		F & 2242 & 4488 & 4498 & 911 & 12139\\
		\hline\hline
	\end{tabular}
	\label{table1}
	\end{center}
\end{table}

\noindent \textbf{Human Detector.}\quad For the human detector model, we utilize all the video frames and their corresponding CSI measurements. Then we split them randomly, in which 9663 data pairs are used for training and 1024 data pairs are used for testing.

\noindent \textbf{Human Segmentor.}\quad If the human in the video is not moving, we cannot leverage the motion vector to generate a reliable human mask. Therefore, we select those video frames that can generate valid motion vector and their corresponding CSI measurements. Then we split them randomly to train the human segmentor model first, in which 8574 data pairs are used for training and 870 data pairs are used for testing.

\noindent \textbf{Forgery Detector.}\quad After the human segmentor model is well trained, we feed the CSI measurements from the human segmentor's dataset into the human segmentor model to prepare the dataset for the forgery detector model. We get 8574 predicted masks and 870 predicted masks for the training and testing sets of the forgery detector, respectively. Then, we concatenate those predicted masks with the motion vector masks to get the labels: 0 if corresponding, else 1. Finally, for the forgery detector model, 8530 data pairs are used for training and 826 data pairs are used for testing.

\subsection{Implementation Details}
The human detector network, human segmentor network, forgery detector network are all implemented for 20 epochs on the Pytorch framework.

\noindent \textbf{Human Detector.}\quad The learning rate starts from $1e-6$ and is divided by 10 for each 5 epochs. The $batchsize=16$ and a RMSprop optimizer \cite{tieleman2012lecture} with $weight\ decay=1e-8$, $momentum=0.9$ is leveraged.

\noindent \textbf{Human Segmentor.}\quad The learning rate starts from $1e-3$ and is divided by 10 for each 5 epochs. The $batchsize=32$ and an adam \cite{kingma2014adam} optimizer with $\beta_1=0.9$, $\beta_2=0.999$ and $weight\ decay=1e-5$ is leveraged.

\noindent \textbf{Forgery Detector.}\quad The learning rate starts from $1e-3$ and is divided by 10 for each 5 epochs. The $batchsize=32$ and an adam \cite{kingma2014adam} optimizer with $\beta_1=0.9$, $\beta_2=0.999$ and $weight\ decay=2e-5$ is leveraged.

\begin{figure*}[htbp]
	\centering
	\includegraphics[width=\textwidth]{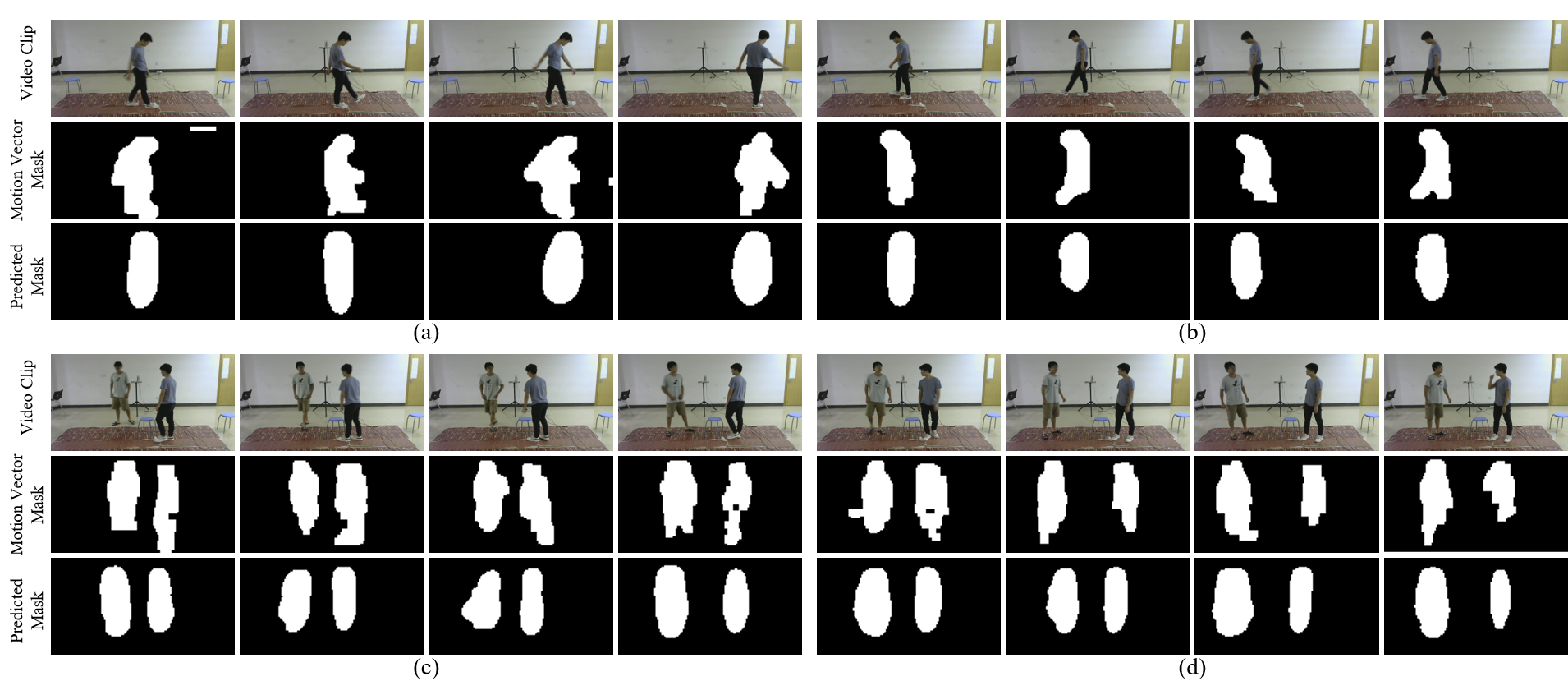}
        \vspace{-0.8cm}
	\caption{Qualitative results of the human segmentor. We provide four video clips in total and each row contains two video clips. In clips (a) and (b), there is one person in the perception field and there are two people in the perception field for clips (c) and (d). For each clip, we show both the motion vector masks and the predicted masks, which reports the predicted masks having the corrected spatial position while the detailed shapes are lost.}
	\label{figure3}
\vspace{-0.5cm}
\end{figure*}

\subsection{Main Results}
In this section, we report the performance of the human detector, human segmentor and the forgery detector. In addition, we also compare the result of the forgery detector with other recent methods.

\noindent \textbf{Qualitative results.}\quad
Since the dataset does not contain manually annotated segmentation labels, we only show the qualitative results of the predicted masks. As shown in Fig.~\ref{figure3}, the predicted masks have the corrected spatial position 
%whether there is one person or two in the perception field
, but the detailed shapes are not recovered. Two reasons can account for the shape difference. First, the spatial resolution of the commercial Wi-Fi signals is less than one decimeter, which makes it difficult for wireless data to capture detailed human boundary information. 
% In addition, it is also an ill-posed problem \cite{ill-posed} to use the 1D Wi-Fi signals to recover a 2D human mask. 
Second, the motion vector only provides coarse supervision compared to masks predicted by neural networks or manual annotation.

\noindent\textbf{Quantitative results.}\quad
We report the quantitative results of the human detector and the forgery detector. In the experiment, we set the number of CSI measurements per video frame $m=5$ and the number of input video frames to the forgery detector $g=7$. As shown in Table~\ref{table2}, the human detector has a perfect performance. This is because the wireless signals are sensitive to moving objects. In the indoor scenario, the motion information can be effectively carried by the Wi-Fi signals: the CSI data contains the spatial information because of the Doppler effect, and as a kind of sampled signal, CSI data contains the temporal information naturally. Therefore, when considering the feature extraction strategy, combining the convolution layer with the LSTM layer is much better than the pure convolution operation as well.

\begin{table}[t]
	\caption{Quantitative results of human detector and forgery detector. We set $m=5$ and $g=7$ for both results.}
	\begin{center}
	\begin{tabular}{cccc}
		\hline\hline
		Module & Acc & FPR & TPR \\
		\hline
		Human Detector & 100\% & 0\% & 100\% \\
		Forgery Detector & 94.38\% & 4.01\% & 92.27\% \\
		\hline\hline
	\end{tabular}
	\label{table2}
	\end{center}
        \vspace{-0.3cm}
\end{table}

The forgery detector also shows its promising performance, whose overall accuracy can reach 94.38\%. We compare our method with other recent approaches \cite{fadl2021cnn, lakshmanan2019surfi, 9089005, huang2021crossmodal}, which are supervised learning-based method. As shown in Table~\ref{table3}, event- and frame-based forgery detection generally have a higher accuracy than those at the object level. However, forgery detection only at the event and frame level will be limited in some situations (i.e., the forgery attack on the entire video).  When considering the working pattern, the overall accuracy of Secure-Mask is only 0.52\% lower compared to its counterpart, Secure-Pose, which needs work in a supervised way. However, our self-supervised approach enables the system to maintain its performance as the environment changes.

\begin{table}[tbp]
    \caption{Comparison result of forgery attack detection methods. Those methods use different datasets to instruct the forgery attack detection at different levels. F: frame-based forgery detection. E: event-based forgery detection. O: object-based forgery detection}
    \begin{center}
    \begin{tabular}{cccc}
        \hline\hline
        Method & Level & Multi-Modal & Acc \\
        \hline
        Fadl \cite{fadl2021cnn} & F &  & 98.0\% \\
        Lakshmanan \cite{lakshmanan2019surfi} & E & \checkmark & 98.9\% \\
        \hline
        Aloraini \cite{9089005} & O &  & 93.18\% \\
        Huang \cite{huang2021crossmodal} (Secure-Pose) & O & \checkmark & 94.9\% \\
        Secure-Mask (Ours) & O & \checkmark & 94.38\% \\ 
        \hline\hline
    \end{tabular}
    \label{table3}
    \end{center}
    \vspace{-0.6cm}
\end{table}

We also report the inference speed of the proposed system. As shown in Table~\ref{table4}, even using a normal GPU, our system can perform in a real-time manner.

\subsection{Ablation Study}

In this section, we conduct extensive ablation experiments to study the core factors of our method. 
% Temporal information is an important for both CSI measurements and video sequences. For the forgery detector, the key is to find a manner that can extract and learn the temporal information effectively. We study the importance of temporal information to forgery detector from two aspects: the number of CSI measurements per video frame $m$ that affects the single predicted video frame, and the number of the input video frames $g$ of the forgery detector, which affects the forgery detector directly.

\noindent \textbf{CSI measurements per video frame $m$.}\quad We first train the human segmentor with different amounts of the CSI measurements per video frame to evaluate the influence of the single predicted video frame on the forgery detector. As reported in Table~\ref{table5}, with the number of the CSI measurements per video frame varying from 1 to 5, the Accuracy increases from 90.25\% to 94.38\%. 
%The working frequency of the wireless device is always higher than the surveillance camera. Thereby, during the sampling procedure, we can assign multiple wireless frames to one video frame. 
This result suggests that an additional temporal information can make it easier for the human segmentor to learn how to decode human motion information from CSI measurements. Indeed, this kind of improvement makes the performance of the forgery detector better.

\begin{table}[t]
	\caption{Inference speed. All results are measured on single NVIDIA 2070 GPU. HD: Human Detector. HS: Human Segmentor. FD: Forgery Detector.}
	\begin{center}
	\begin{tabular}{cccc}
		\hline\hline
		Module & HD & HS & FD \\
		\hline
		FPS & 230 & 70 & 280 \\
		\hline\hline
	\end{tabular}
	\label{table4}
	\end{center}
\end{table}

\begin{table}[t]
\vspace{-0.5cm}
	\caption{\textbf{CSI measurements per video frame.}\quad The performance improve as the amount of the CSI measurements per video frame increases. The Ratio here refers to the CSI measurements per video frame.}
	\begin{center}
    \begin{tabular}{cccc}
        \hline\hline
        Ratio ($m$) & Acc & FPR & TPR \\
        \hline
        1 & 90.25\% & 10.11\% & 90.57\% \\
        3 & 91.25\% & 3.47\% & 84.60\% \\
        5 & 94.38\% & 4.01\% & 92.27\% \\
        \hline\hline
    \end{tabular}
    \label{table5}
    \end{center}
\vspace{-0.5cm}
\end{table}

\noindent \textbf{Number of the input video frames to forgery detector $g$.}\quad We then train the forgery detector with different numbers of the input video frames to evaluate the effect of the amount of the video frames on the model. As reported in Table~\ref{table6}, with the number varying from 3 to 7, the accuracy increases from 87.17\% to 94.38\%. This result shows that feeding more video frames into the model does improve its performance. The LSTM component in the forgery detector model can account for this since the LSTM component can learn the forgery information more efficiently when additional special information combined with temporal information is provided. As the video was recorded at 7.5 FPS from the camera, we set the largest amount of the input video frames as 7 to avoid fail detection with the short-time forgery. Moreover, we argue that if we recorded the video at a higher FPS, the forgery detector can achieve even better performance.

\begin{table}[tbp]
    \caption{\textbf{Number of the input video frames.}\quad The performance improve as the number of the input video frames increases}
    \begin{center}
    \begin{tabular}{cccc}
        \hline\hline
        Frames ($g$) & Acc & FPR & TPR \\
        \hline
        3 & 87.17\% & 4.35\% & 79.08\% \\
        5 & 92.01\% & 4.39\% & 87.50\% \\
        7 & 94.38\% & 4.01\% & 92.27\% \\
        \hline\hline
    \end{tabular}
    \label{table6}
    \end{center}
\vspace{-0.5cm}
\end{table}

\section{Conclusion}

In this paper, we build a novel self-supervised system for video forgery detection eliminating the need for external annotations. Notably, our method achieves comparable performance against the previous supervised methods in forgery detection.

\bibliographystyle{IEEEbib}
\bibliography{references}

\end{document}